\documentclass[preprint]{aastex}
\usepackage{graphicx}
\usepackage{natbib}
\usepackage{color}

\shorttitle{The 17~GHz active region number}
\shortauthors{Selhorst et al.}

\begin{document}

\title{The 17~GHz active region number}

\author{C. L. Selhorst }
\affil{IP\&D - Universidade do Vale do Para\'iba - UNIVAP, S\~ao Jos\'e dos Campos, Brazil }
\email{caius@univap.br}
\author{J. E. R. Costa}
\affil{CEA, Instituto Nacional de Pesquisas Espaciais, S\~ao Jos\'e dos Campos, Brazil}
\author{C. G. Gim\'enez de Castro }
\affil{CRAAM, Universidade Presbiteriana Mackenzie, S\~ao Paulo, Brazil \\
and IAFE, Universidad de Buenos Aires/CONICET, Buenos Aires, Argentina}
\author{A. Valio}
\affil{CRAAM, Universidade Presbiteriana Mackenzie, S\~ao Paulo, Brazil}
\author{A. A. Pacini }
\affil{IP\&D - Universidade do Vale do Para\'iba - UNIVAP, S\~ao Jos\'e dos Campos, Brazil }
\and 
\author{K. Shibasaki}
\affil{Nobeyama Solar Radio Observatory /NAOJ,  Minamisaku, Nagano 384-1305, Japan}

\begin{abstract}
{We report the statistics of the number of active regions  ($NAR$) observed at 17~GHz with the Nobeyama Radioheliograph between 1992, near the maximum of cycle 22, and 2013, that also includes the maximum of cycle 24, and we compare with other activity indexes. We find that $NAR$ minima are shorter than those of the sunspot number ($SSN$) and radio flux at 10.7~cm ($F10.7$).  This shorter $NAR$ minima could reflect the presence of active regions generated by faint magnetic fields or spotless regions, which were a considerable fraction of the counted active regions.
 The ratio between the solar radio indexes $F10.7/NAR$ shows
a similar reduction during the two minima analyzed, which contrasts with the increase of the ratio of both radio indexes
in relation to the $SSN$ during the minimum of cycle 23-24.  These results indicate that the radio indexes are more sensitive to weaker magnetic fields than those necessary to form sunspots, of the order of 1500~G. The analysis of the monthly averages of the active region brightness temperatures shows that its long term variation  mimics the solar cycle, although, due to the gyro-resonance emission, a great number of intense {\em spikes} are observed  in the maximum temperature study. The decrease, in number, of these spikes is also evident during the current cycle 24, a consequence of the sunspot magnetic field weakening in the last years.} 
\end{abstract}
 

\maketitle

\section{Introduction}

The unusual solar activity low level, inferred by different indexes during cycle 24, launched new discussions concerning the future behavior of the Sun. The oldest and most famous solar activity index is the counting of dark regions on the solar surface, consisting in the so called Sunspot Number ($SSN$). A sunspot appear because of the obstruction of convective movement that occurs at the base of the photosphere due to the existence of a strong magnetic field in the region. Thus, the $SSN$ index is associated only with photospheric features of the solar magnetic variability. 

\cite{Penn2006} determined that the minimum magnetic field intensity necessary for sunspot formation is 1500~G and also suggested that there is, in average, a linear decrease in the sunspot maximum magnetic field intensity occurring along the cycles. This reduction was confirmed in a recent study, which shows that sunspots maximum magnetic field intensities in the period of 1998-2002 have a normal distribution with a mean of $2436\pm26$~G and a width of $323\pm20$~G, while during the period 2008-2011 the mean value dropped to $1999\pm13$~G and a width of $276\pm9$~G \citep{Livingston2012}. An alternative interpretation, proposed by \cite{Nagovitsyn2012}, suggested that the mean magnetic field reduction could be explained by a gradual decrease in the number of large sunspots, whereas the number of small ones increased.

Another classical index used to measure the solar activity is the 10.7~cm daily mean radio flux (hereafter referred as $F10.7$), which is generated at coronal heights and related to the presence of active regions and the occurrence of flares. During cycle 24, $F10.7$ also showed a decrease compared to previous cycles. However, its reduction rate is different from the $SSN$ decrease, destroying the linear relation traditionally assumed to exist between these two solar indexes, and making the prediction of $SSN$ based on $F10.7$ to be overestimated in the last two cycles \citep{Hathaway2010,Livingston2012}.

Since the radio emission of non-flaring active regions in the cm range depends on free-free radiation from the hot plasma trapped in the magnetic field loops and from the gyro-resonance radiation due to strong magnetic fields, the results obtained by \cite{Livingston2012} 
suggests that the presence of magnetic fields is still contributing to $F10.7$, i.e., the magnetized regions are still emitting at radio frequencies although the magnetic field is not strong enough to generate a sunspot.     
 
At 17~GHz, the gyro-resonance emission from active regions overcomes the free-free radiation when the third harmonic (associated to magnetic field intensities of $2000$~G) is produced in the transition region or at coronal heights \citep{Shibasaki2011}, which results in high brightness temperature: above ten times the quiet Sun values \cite[$T_B\gtrsim10^5$~K, ][]{Gelfreikh1999}. Thus, the active region maximum brightness temperatures observed at 17~GHz should reflect the magnetic field reduction reported by \cite{Livingston2012}.

To investigate the relationship between $SSN$ and $F10.7$, we report in this work a statistical analysis of the number of active regions (hereafter referred to as $NAR$) measured in the 17~GHz maps obtained since 1992 by the Nobeyama Radioheliograph  \cite[NoRH, ][]{Nakajima1994}. 

\section{Data Analysis and Results}
 
The Nobeyama Radioheliograph has been obtaining daily solar maps at  17~GHz with 15~arcsec resolution since 1992. Its database provides an unprecedented opportunity to perform statistical studies at radio frequencies, such as the analysis of the variation of solar radius and polar brightening along the solar cycle \citep{Selhorst2011,Gopal2012}. In this work, we calculate the daily number of active regions observed in Nobeyama 17~GHz maps  and compare the results with the sunspot number ($SSN$)  and the 10.7~cm radio flux ($F10.7$).   
 
 \begin{figure}[!h]
\centerline{ {\includegraphics[width=16.cm]{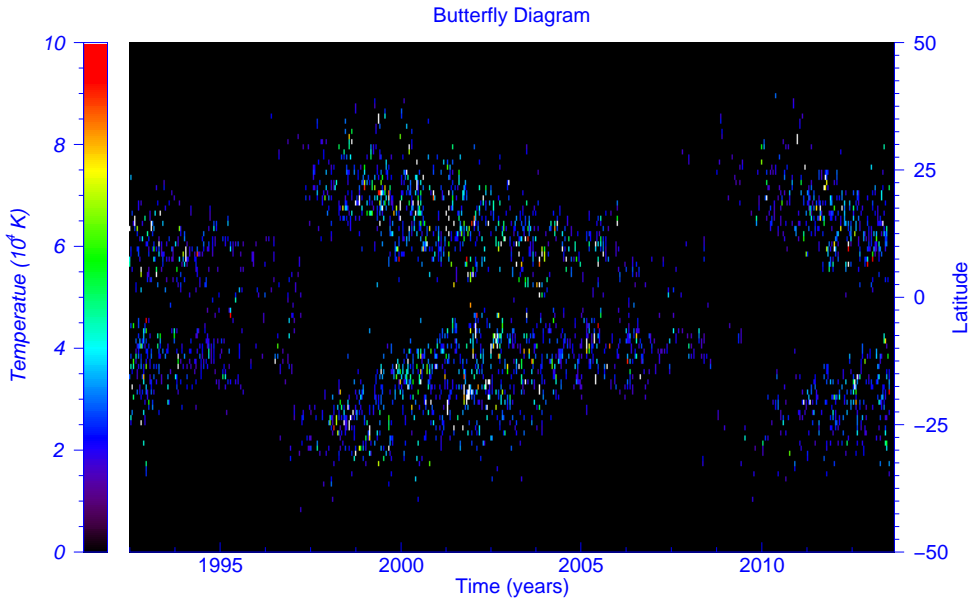}}}
\caption{Latitude positions of the 17~GHz  active regions detected in the analysis. The color scale represents the active region maximum brightness temperature ($T_B<1\times10^5$~K).}
\label{fig:bf}
\end{figure}
 
The identification of each active region in the 17~GHz maps was automatized using computerized procedures. To avoid small bright features dispersed over the solar disk, as week as the polar bright patches, the following criteria were adopted:
\begin{itemize}
\item Size greater than 150~$pixels^2$ ($\sim300$ MSS (millionth of solar surface)); 
\item Latitudes between $-45^\circ$ and $+45^\circ$;
\item Maximum brightness temperature ($T_{B_{max}}$) at least 40\% greater than the quiet Sun value.
\item Discard active regions at the limb.
\end{itemize}   

For the studied period (1992-2013) almost 7500 maps were analyzed, in which over 16000 active regions were detected by the method described above. During this period, only five false active regions generated by prominence eruption \citep[PE, ][]{Gopal2013} were detected and removed by visual inspection, none of them occurred during minima periods. This low count is coherent with the statistic of 1.4 PEs per month during the rise phase of the 22-23 minimum, and of 0.5 PEs for the 23-24 minimum \citep{Gopal2012}. Moreover, the possible contamination will became statically negligible, since the large scale brightening caused by the PEs are shorter than the interval between the daily NoRH maps used here, which were all obtained at the same time (local midday). 

Figure~\ref{fig:bf} shows the latitude positions of the 17~GHz active regions as a function of time,  where the color scale represents the maximum brightness temperature, that saturates at $T_B=1\times10^5$~K. As expected, the $NAR$ latitude distribution follows the butterfly diagram. The majority of active regions are cooler than  $4\times10^4$~K, that is compatible with bremsstrahlung as the main mechanism emission \cite{Selhorst2008}.   
 
 \begin{figure}[!h]
\centerline{ {\includegraphics[width=11.cm]{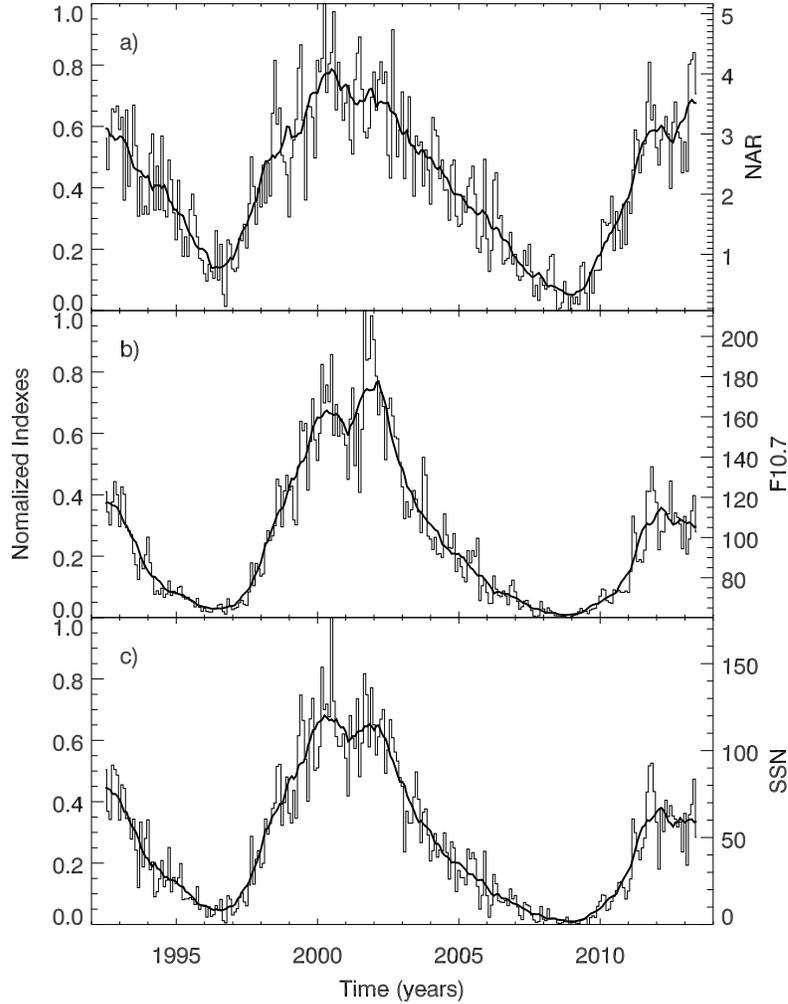}}}
\caption{Monthly means of {\bf a)}  the number active regions observed at 17~GHz, {\bf b)} the 10.7~cm radio flux, and {\bf c)} the sunspot number. The values on the left side are normalized ones, while the original values are displayed on the right hand side scale. The thick line represents a running mean of 1 year (12 points).}
\label{fig:numbers}
\end{figure}
 
The $NAR$ monthly average is plotted on Figure~\ref{fig:numbers}a, along with $F10.7$ and $SSN$ monthly means 
(Figures~\ref{fig:numbers}b and \ref{fig:numbers}c, respectively). Although the $NAR$ also mimics the solar cycle, one can see small discrepancies between the number of active region and the classical solar index plotted on Figure~\ref{fig:numbers}, in which the thick line represents a running mean of 1 year (12 points).

\begin{figure}[!h]
\centerline{ {\includegraphics[width=11.cm]{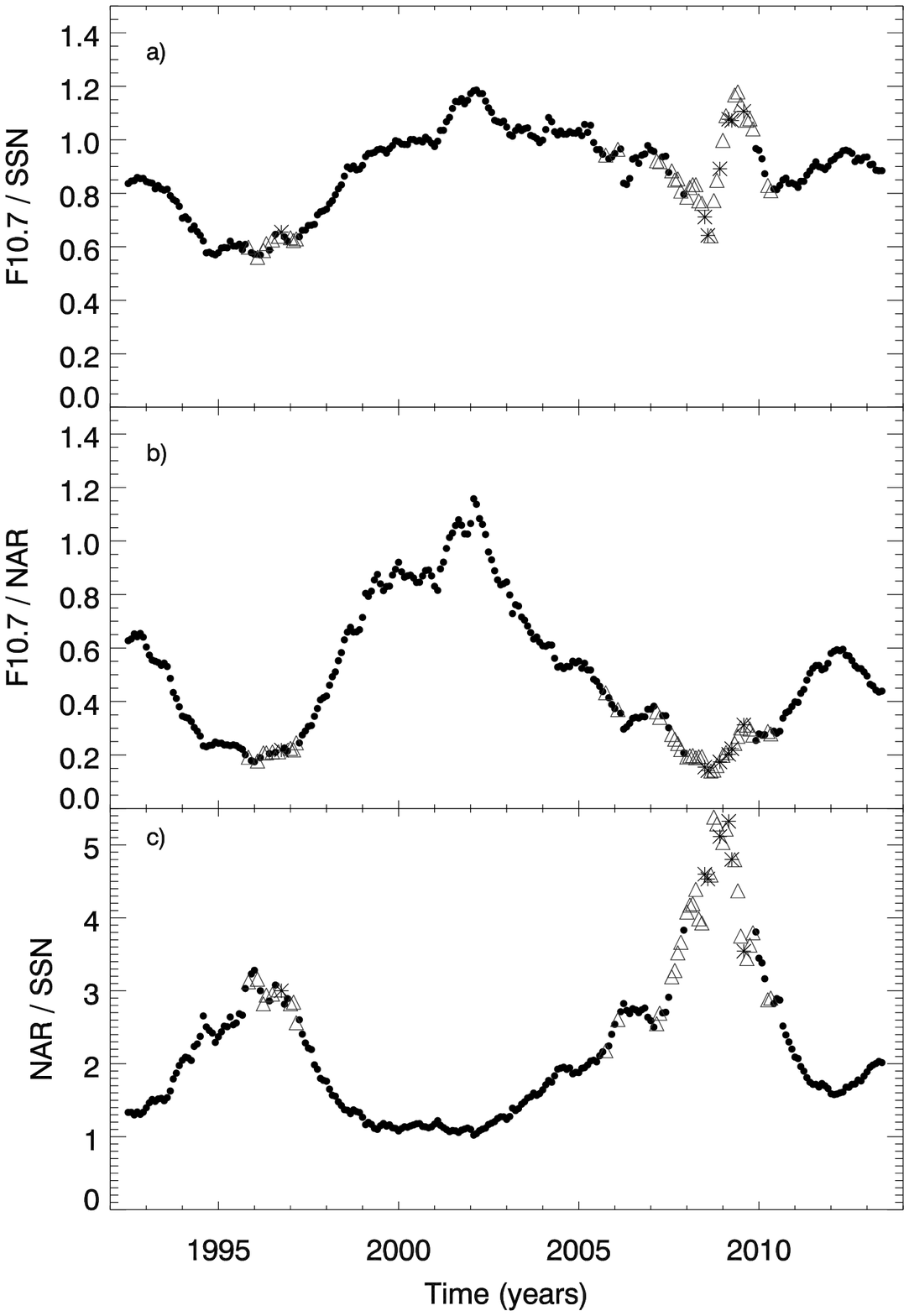}}}
\caption{The temporal variation of the ratio between the solar index:  a) $F10.7/SSN$, b) $F10.7/NAR$, and c) $NAR/SSN$. 
The points represent the months with $\rm SSN>10$, whereas the triangles and asterisks are plotted when $\rm 1\leq SSN\leq10$ and $\rm 0\leq SSN<1$, respectively.}
\label{fig:ratio}
\end{figure}

To quantify these differences, ratios between the yearly running mean of the three indexes are plotted on 
Figure~\ref{fig:ratio}. In all panels, the dots were used for the months with $\rm SSN>10$, whereas the triangles 
and asterisks represent the months with $\rm 1\leq SSN\leq10$ and $\rm 0\leq SSN<1$ respectively, that is, 
minimum activity. The ratio between $F10.7$ and $SSN$ (Figure~\ref{fig:ratio}a) shows distinct behavior during 
the two minima analyzed. This ratio is reduced during the cycle minimum 22-23, but increases during the  quietest months of cycle  23-24 minimum. On the other hand, the ratio $F10.7/NAR$ shows a similar reduction during both minima (Figure~\ref{fig:ratio}b). The ratio between the yearly $NAR$ and  $SSN$ running means, plotted on Figure~\ref{fig:ratio}c, shows an increase during both minima intervals, and reaches its highest level during the prolonged minimum between the cycles 23 and 24 (2008-2010). 

The two minima of the radio indexes ratio ($F10.7/NAR$) are similar, whereas the minima of the ratios $F10.7/SSN$ and $NAR/SSN$  clearly change. These differences can be due to the great reduction observed in the $SSN$ during the prolonged last minimum. For example, during minimum 22-23 there was only one month with a monthly  average $SSN$ less than 1, while during minimum 23-24, five months had $SSN<1$ (Figure~\ref{fig:numbers}). Moreover, there are a small time lag between the ratios $F10.7/SSN$ and $NAR/SSN$ (Figures~\ref{fig:ratio}a and \ref{fig:ratio}c, respectively), which could be caused by the phase shift in the north-south SSN maxima observed \cite[see e.g., ][]{Hathaway2010}, that is  indistinguishable in the total solar flux measured by F10.7.

For the 6 months with $SSN<1$, $F10.7$ remains almost constant, with a mean value of $\overline{F10.7}=61.6\pm0.9 \rm{~Wm^{-2}Hz^{-1}}$  and the monthly  $NAR$ never drops to zero. Moreover, the presence of spotless active regions \cite[see e.g., ][]{Gary1990} were commonly observed during these low activity months. During the quietest period observed (2008-2009) there were 329 active regions counted  and 109 ($33\%$) of them were identified on days without sunspots. These results indicate that the $F10.7$ excess with respect to $SSN$ reported for the last cycles could be due to the increase in the relation between $NAR$ and $SSN$ (Figure~\ref{fig:ratio}c). In other words, the active regions responsible for the $F10.7$ during both minima were formed at the same rate, instead of the lack of sunspots. 

Considering that the main emission mechanisms of non-flaring  active regions at 17~GHz are free-free and the gyro-resonance radiation formed at the third harmonic of the gyro-frequency \citep{Selhorst2008,Shibasaki2011}, one can expect a reduction in the maximum brightness temperature observed at 17~GHz following  the reduction of the magnetic field intensity observed in sunspots during the current cycle \citep{Penn2006,Schad2010}. To verify this assumption, monthly averages of the active region mean and maximum brightness temperatures were calculated. To avoid the high brightness temperatures related to flares, and also considering that the gyro-resonance mechanism can increase the temperatures to values close to $5\times 10^5$~K \citep{Selhorst2009}, our analysis was restricted to active regions with $T_{B_{max}}<6\times 10^5$~K, neglecting only around 0.7\% of the total active regions selected. 

\begin{figure}[!h]
\centerline{ {\includegraphics[width=11.cm]{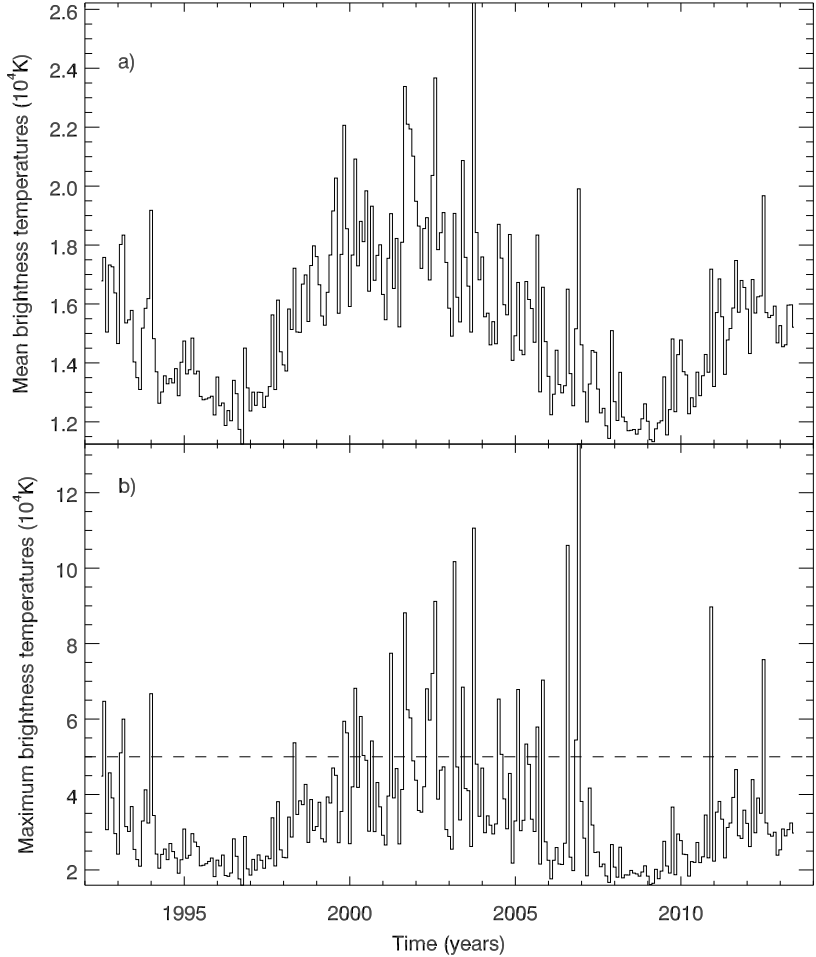}}}
\caption{Monthly average of a) mean and b) maximum brightness temperatures observed at 17~GHz. The dashed horizontal line was plotted at $T_B=5\times 10^4$~K. }
\label{fig:tb}
\end{figure}

The result of the brightness temperature analysis is shown on Figure~\ref{fig:tb}. While the variation of the mean brightness temperatures smoothly mimics  that of solar cycle (Figure~\ref{fig:tb}a), the maximum temperature values show a great number of intense spikes (Figure~\ref{fig:tb}b). Since some active regions can be visible on the solar disk for more than 10 days, the presence of these active regions with gyro-resonance contribution increases the maximum brightness temperature average, which appears as intense spikes above 
the mean values. It is seen in Figure~\ref{fig:tb}, that only during two months of the current cycle  (after 2009) $T_{B_{max}} > 5\times10^4$~K, in contrast with the 27 months of the period between 1998 and 2007. 

Curiously, some intense spikes occurred between 2006 and 2007, when the number of sunspots were very low. Nonetheless, this period coincides with an increase of the relative number of large and intense sunspots \citep{Nagovitsyn2012}. After that, the percentage of large sunspots reduces, while the number of small ones increases \citep{Nagovitsyn2012}.  

\section{Discussion and conclusions}

The Nobeyama Radioheliograph has been daily observing the Sun for over two decades, which makes possible the  statistical study 
presented here. The long solar minimum, observed between cycles 23 and 24, brought some new information concerning the quiescent 
solar behavior, such as the existence of active regions at 17~GHz without photospheric sunspots. This evidence indicates that the lower 
limit of magnetic fields for active region formation at 17~GHz is smaller than that necessary for sunspot formation inferred by \cite{Penn2006}. 

Comparing the $F10.7$ and $SSN$ monthly averages, \cite{Livingston2012} reported that the difference between the predicted and the observed $SSN$ reflects the reduction in the umbra magnetic field intensity observed in the last 20 years. This divergence indicates that the magnetic regions are still there, however, the magnetic fields are not strong enough to generate sunspots. To determine the lower limit of the magnetic field intensity to create an active region at 17~GHz, the ratios $F10.7/SSN$ and $F10.7/NAR$ were calculated throughout the cycle. The ratio between the solar radio indexes $F10.7$ and $NAR$ follows a similar reduction during the two minima analyzed (Figure~\ref{fig:ratio}b). On the other hand, the reduction observed in the ratio $F10.7/SSN$ during the minimum 22-23 is not observed in the prolonged minimum 23-24 (Figure~\ref{fig:ratio}a). This change can be due to the absence of sunspots during long periods of cycle 23-24 minimum, while active regions are detected.  For this reason, the ratio $NAR/SSN$ increased (Figure~\ref{fig:ratio}c), i.e., active regions were formed with lower number of  sunspots, and occasionally without them \citep{Gary1990}. Actually, the presence of spotless regions were quite common during the prolonged  minimum, for example, 33\% of active regions counted in 2008-2009 were observed on days without sunspots. Furthermore, according to our results, it is also possible to deduce that the minimum photospheric magnetic field intensity necessary to form the 17~GHz active regions  is $<1500$~G, which  is the minimum to form sunspots.

Complementary to the active region magnetic field measurements, the analysis of active region maximum brightness temperatures clearly reflects the magnetic field intensity decrease reported by \cite{Livingston2012}, which could be due to a decrease in the number of large and intense sunspots, accompanied by an increase in the number of small ones \citep{Nagovitsyn2012}. At 17~GHz the active region brightness temperatures can reach values of $T_B>10^5$~K if the third harmonic of gyro-resonance (2000~G) occurs above the transition region \citep{Gelfreikh1999}. The occurrence of these intense active active regions was strongly reduced in the current maximum, which is reflected in the low number of months with mean brightness temperatures above $5\times10^4$~K shown in Figure~\ref{fig:tb}b. 

Taking into account that gyro-resonance emission at 17~GHz needs a minimum magnetic field intensity of 2200~G at  photospheric levels \citep{Vourlidas2006}, if the predictions made by \cite{Livingston2012} for the magnetic field 
intensity for the period of 2012-2016  are confirmed, it is probably that hot active regions ($T_B>10^5$~K) 
will almost disappear from the solar maps at 17~GHz. 

\cite{Henney2012} propose a method to forecast $F10.7$ estimating empirically the future global photospheric magnetic field using the flux transport model ADAPT. Nonetheless, this method has many unknowns, mostly derived from the lack of spatial observations at 10.7~cm wavelenghts. Since we have shown in this paper that $F10.7$ and $NAR$ correlates well, the Nobeyama 17~GHz daily maps may be used  to better constraint the forecasting method proposed by \cite{Henney2012}, as an alternative to the lack of solar maps at 10.7~cm, with the advantage that at 17~GHz  we have good spatial  information and a 20 years long database. 

\acknowledgments

We would like to thank the Nobeyama Radioheliograph, which is
operated by the NAOJ/Nobeyama Solar Radio Observatory.  C.L.S. and A.A.P. acknowledge financial 
support from the S\~ao Paulo Research Foundation (FAPESP), grant number 2012/08445-9.

\end{document}